\documentstyle[prl,preprint,eqsecnum,aps,epsf]{revtex}
\newcommand{\beq}{\begin{equation}}
\newcommand{\eeq}{\end{equation}}
\newcommand{\beqs}{\begin{eqnarray}}
\newcommand{\eeqs}{\end{eqnarray}}
\newcommand{\nn}{\nonumber}
\newcommand{\dd}[2]{\frac {\partial #1}{\partial #2}}

\draft
\begin{document}
\preprint{GTP-97-02}
\title
{Propagators for spinless and spin-1/2 Aharonov-Bohm-Coulomb systems}
\author
{D.K.Park$^a$ and Sahng-Kyoon Yoo$^b$}
\address
{$^a$ Department of Physics, Kyungnam University, Masan 631-701, Korea \\
$^b$ Department of Physics, Seonam University, Namwon, Chunbuk 590-170, Korea}
\date{\today}
\maketitle
\begin{abstract}
\indent The propagator of the spinless Aharonov-Bohm-Coulomb system is 
derived by following the Duru-Kleinert method. 
We use this propagator to explore the spin-1/2 Aharonov-Bohm-Coulomb 
system which
contains a point interaction as a Zeeman term. Incorporation of the 
self-adjoint extension method into the Green's function formalism 
properly allows us to derive the finite propagator of the spin-1/2 
Aharonov-Bohm-Coulomb system. As a by-product, the relation
between the self-adjoint extension parameter and the bare coupling
constant is obtained. Bound-state energy spectra of both spinless and spin-1/2 
Aharonov-Bohm-Coulomb systems are examined.

\end{abstract}

\pacs{}

\section{Introduction}
A great deal of attention of the 
Aharonov-Bohm(AB) effect\cite{aha59}, which sheds light on the non-trivial 
physical significance of the scalar and vector potentials at quantum level, 
has been paid
in recent years in the context of anyonic\cite{wil82}, 
cosmic string\cite{alf89}, and (2+1)D gravity theories\cite{des84}.
Since anyon, a two-dimensional object, carries the magnetic flux\cite{hag84},
the dominant interaction between anyons is of AB type. Arovas {\it et al}
have pointed out in their seminal paper\cite{aro85} 
by calculating the second virial coefficient 
that the statistics of anyon system interpolates 
between bosons and fermions, which is the most important property of 
anyon system for attacking the 
high-$T_c$ phenomena in superconductivity.

Many authors investigated whether or not this property can be
maintained when the
spin degree of freedom is included. One remarkable difference of the spin-1/2
AB problem from the spinless case is a point interaction potential term
which occurs as a mathematical description of the Zeeman interaction of spin
with a magnetic flux tube.
Gerbert dealt with this problem\cite{ger89} by applying the self-adjoint 
extension method\cite{alb88} to the partial wave. He claimed that when
$\mid m + \alpha \mid < 1$, where $m$ and $\alpha$ are angular momentum
quantum number and flux parameter respectively, both regular and 
irregular radial solutions at the origin are allowed. Being compatible
with the boundary condition derived by self-adjoint extension technique, 
his solution naturally contains undetermined real parameter, say $\theta$,
which is called as self-adjoint extension parameter.

Hagen also analyzed the same problem on the physical 
ground\cite{hag90}. He chose the physically motivated expression
of the flux tube
\begin{eqnarray*}
H \propto \lim_{R \rightarrow 0} \frac{1}{R} \delta(R - r)
\end{eqnarray*}
where $H$ is a magnetic field, and solved the radial Schr\H{o}dinger 
equation at the
$r < R$ and $r > R$ regions separately. Upon applying the matching 
conditions at $r = R$ in the $R \rightarrow 0$ limit, he argued that 
when both $\mid m + \alpha \mid < 1$ and $\mid m \mid + \mid m + \alpha \mid = -
\alpha s$ 
are satisfied simultaneously, where $s$ is twice the spin quantum
number, the physically relevant wave is the irregular solution at the origin.
Although the self-adjoint extension method does not yield the latter one, 
it is easily shown that Hagen's result coincides with Gerbert's when the 
self-adjoint extension parameter $\theta$ equals $\pi / 4$. Hagen also
calculated\cite{blu90} the second virial coefficient for the spin-1/2 
AB system and showed that it is completely different from that for the spinless
AB system.

A series of the above-mentioned papers on the spin-1/2 AB problem raises an 
issue on the physical meaning of the self-adjoint extension parameter.
Some authors\cite{cou93} tried to dig out its physical meaning  
in the internal
structure of the magnetic flux tube. This kind of approach
, however, encountered with the criticism\cite{hag93}  
that their calculation within the framework of the Dirac equation
tends not to be reliable because of the occurrence of Klein's paradox.
Jackiw\cite{jac91} also approached the same issue from the different
point of view. Analyzing the two- and three-dimensional delta-function potentials,
he asserted that the self-adjoint extension parameter has a certain relation 
with a renormalized(or bare) coupling constant.

Recently his result was used for the derivation of propagator in  
the spin-1/2 AB 
system by one of us\cite{park95}. In Ref.[14] it is shown that 
the relation between the self-adjoint extension parameter and the 
renormalized coupling constant is more consistently derived by incorporating
the self-adjoint extension method into the Green's function formalism
properly. The method suggested in Ref.[14] is also applied to the  
one-dimensional $\delta'$-potential case\cite{park96}.

Since the AB interaction is accomplished between charge and magnetic flux, 
the interacting anyon system naturally involves the Coulomb modification. The 
effect of Coulomb potential in the spin-1/2 AB problem is discussed by Hagen
in Ref.[12] by solving the Schr\H{o}dinger equation. Since the Feynman
propagator\cite{fey65} is essential and useful for analyzing the 
time-dependent scattering and statistical property of anyon system, it is 
very important to derive the propagator of the Aharonov-Bohm-Coulomb(ABC)
system. In the present paper we will derive the propagators of the spinless and 
spin-1/2 ABC systems by using the 
Duru-Kleinert pseudotime method\cite{duru79} and the essential idea in Ref.[14].

The non-relativistic solution of hydrogen atom has been 
a long-standing problem
in the path-integral framework. Duru and Kleinert solved 
the problem with a help of Kustaanheimo-Stiefel(KS)\cite{kus65}
transformation and through the introduction of the dimensional extension 
and pseudotime.
Their essential idea is based on the following observations: if the 
system possesses a Feynman path-integral for the time evolution amplitude,
it does so also for the fixed-energy amplitude $K[\vec{x}_b, \vec{x}_a ; E]$.
Their idea came up with the following time-sliced formula
\beqs
& &K[\vec{x}_b, \vec{x}_a ; E] \\ \nn
&=& \lim_{N \rightarrow \infty} (N+1) \int_{0}^{\infty} d \epsilon_s
f_r(\vec{x}_b) f_l(\vec{x}_a)
\int \left( \prod_{j=1}^{N} d \vec{x}_j \right)
\int \left( \prod_{j=1}^{N+1} \frac{d \vec{p}_j}{(2 \pi)^D} \right)
e^{i A_E^N}
\label{fixeden-11}
\eeqs 
where
\beq
A_E^N = \sum_{j=1}^{N+1}
        \left[
              \vec{p}_j \cdot (\vec{x}_j - \vec{x}_{j-1}) - \epsilon_s
              f_l(\vec{x}_j) \left\{
                                   H(\vec{p}_j, \vec{x}_j) - E \right\}
              f_r(\vec{x}_{j-1})
                                       \right].
\label{action-12}
\eeq
Here, $H$ and $E$ are the Hamiltonian of a given system and its eigenvalue
respectively, and $f_l(\vec{x})$ and $f_r(\vec{x})$ regulating functions
defined in chapter 12 of Ref.[19].
Pseudotime $s$ is defined as $ds/dt = f_l(\vec{x}) f_r(\vec{x})$ and 
$(N+1) \epsilon_s = s_b - s_a \equiv s$.

In this paper the propagators of the spinless and spin-1/2
ABC systems are derived. Sec.II gives the derivation of  
the propagator of the spinless
ABC system using Eq.(1.1) and 
{\it Levi-Civit\`{a}} transformation that is a two-dimensional version 
of the KS transformation. In Sec.III we will review Ref.[14] briefly
to prepare the calculation in spin-1/2 ABC system.
In Sec.IV the derivation of the propagator of the spin-1/2 ABC system 
is presented by the appropriate incorporation of 
the self-adjoint extension method into the Green's function formalism.
A brief conclusion is given in  Sec.V.
Throughout this paper, we take $\hbar =1$ for simplicity.
\newpage

\section{Propagator for spinless ABC system}

In this section we will derive the propagator for a spinless ABC system 
by following the method used by Kleinert[17, 19]. Let us start with the 
Hamiltonian
\beq
H = \frac{(\vec{p} - e \vec{A})^2}{2M} + \frac{\xi}{r}
\label{hamil-21}
\eeq
where the AB potential is 
\beq
\vec{A} = \frac{\alpha}{e} \epsilon_{ij} \frac{r_j}{r^2}
\label{vetpo-22}
\eeq
in Coulomb gauge and $\epsilon_{12} = 1$.

Then the momentum integration of Eq.(1.1) can be performed straightforwardly
and the resultant fixed-energy amplitude is
\beqs
& &K[\vec{x}_b, \vec{x}_a ; E]  \\  \nn
&=& (N+1) \int_0^{\infty} d \epsilon_s f_r(\vec{x}_b) f_l(\vec{x}_a)
  \left( 
        \frac{M}{2 \pi i \epsilon_s f_r(\vec{x}_a) f_l(\vec{x}_b)}
                                              \right)^{\frac{1}{2}} 
 \int
 \left[ \prod_{j=1}^{N} 
        \frac{d \vec{x}_j}{(2 \pi i \epsilon_s f_l(\vec{x}_j) f_r(\vec{x}_j) /
              M)^{1/2}}
                                                      \right]  \\  \nn
&\times& exp i \sum_{j=1}^{N+1}
       \Bigg[ \frac{M}{2 \epsilon_s f_l(\vec{x}_j) f_r(\vec{x}_{j-1})}
              (\vec{x}_j - \vec{x}_{j-1})^2 + e \vec{A}_j \cdot 
               (\vec{x}_j - \vec{x}_{j-1})  \\  \nn
& &
\hspace{7.0cm}
             - \epsilon_s f_l(\vec{x}_j)
               \left( \frac{\xi}{r_j} - E \right) f_r(\vec{x}_{j-1})
                                                                    \Bigg].
\label{fixeden-23}
\eeqs
In deriving Eq.(2.3), we fixed $D=2$.

With the following choice 
\beqs
f_l(\vec{x})&=& r^{1 - \lambda}  \\  \nn
f_r(\vec{x})&=& r^{\lambda}
\label{regulat-24}
\eeqs
$K[\vec{x}_b, \vec{x}_a ; E]$ can be simplified in a form
\beq
K[\vec{x}_b, \vec{x}_a ; E]  
=(N+1) \int_0^{\infty} d \epsilon_s 
   \frac{M}{2 \pi i \epsilon_s} 
   \left( \frac{r_a}{r_b} \right)^{1 - 2 \lambda}
   \int \left[ 
              \prod_{j=2}^{N+1} \frac{M}{2 \pi i \epsilon_s r_{j-1}}
                                                 d \Delta \vec{x}_j
                                                          \right]
   e^{i S_{0,E}^N}
\label{fixeden-25}
\eeq
where
\beqs
S_{0,E}^N&=& \sum_{j=1}^{N+1} 
           \Bigg[ \frac{M}{2 \epsilon_s r_j^{1 - \lambda} r_{j-1}^{\lambda}}
                  (\vec{x}_j - \vec{x}_{j-1})^2 + e \vec{A}_j \cdot
                  (\vec{x}_j - \vec{x}_{j-1}) \\  \nn
& &
\hspace{5.0cm}
                - \xi \epsilon_s 
                  \left( \frac{r_{j-1}}{r_j} \right)^{\lambda} + 
                  \epsilon_s E r_j \left( \frac{r_{j-1}}{r_j} \right)^{\lambda}
                                               \Bigg]
\label{action-26}
\eeqs
whose continuum limit is 
\beq
S_{0,E}[\vec{x}, \vec{x}'] = -\xi s + 
\int_0^s ds \left[
                  \frac{M}{2r} \vec{x}'^2 + e \vec{A} \cdot \vec{x}' + E r
                                                           \right]
\label{action-27}
\eeq
and $\vec{x}' = d \vec{x} / ds$. Since continuum limit of 
$K[\vec{x}_b, \vec{x}_a ; E]$ is independent of $\lambda$\cite{kle95}, we 
set $\lambda = 1/2$, which yields 
\beq
K[\vec{x}_b, \vec{x}_a ; E]  
=(N+1) \int_0^{\infty} d \epsilon_s
         \frac{M}{2 \pi i \epsilon_s}
         \int \left[
                    \prod_{j=2}^{N+1} \frac{M}{2 \pi i \epsilon_s r_{j-1}}
                    d \Delta \vec{x}_j
                                              \right]
         e^{i S_{0,E;\lambda = 1/2}^N}.
\label{fixeden-28}
\eeq

We now apply the {\it Levi-Civit\`{a}} transformation
\beq
\left( \begin{array}{c}
        x \\ y
       \end{array} \right)
       = A(\vec{u})
\left( \begin{array}{c}
       u^1 \\ u^2
       \end{array} \right),
\label{LC-29}
\eeq
where
\beq
A(\vec{u})=
\left( \begin{array}{cc}
   u^1 &  -u^2 \\
   u^2 &   u^1
  \end{array} \right),
\label{LC-210}
\eeq
to the path-integral calculation of $K[\vec{x}_b, \vec{x}_a ; E]$.
>From the definition of {\it Levi-Civit\`{a}} transformation (\ref{LC-29}) and
(\ref{LC-210}) it is easily shown that
\beqs
x^2 + y^2&=& \left[(u^1)^2 + (u^2)^2 \right]^2  \\ \nn
\Delta x^2 + \Delta y^2&=& 4 \left[(u^1)^2 + (u^2)^2 \right] 
                     \left[ (\Delta u^1)^2 + (\Delta u^2)^2 \right] \\ \nn
\frac{\partial (x, y)}{\partial (u^1, u^1)}&=& 2^2 r.
\label{LC-211}
\eeqs
Furthermore, the {\it Levi-Civit\`{a}} transformation of the AB potential term
is
\beqs
\vec{A}_j \cdot d \vec{x}_j
&\equiv& \alpha \frac{y_j \Delta x_j - x_j \Delta y_j}{r_j^2} \\ \nn
&=& 2 \alpha \frac{u^2_j \Delta u^1_j - u^1_j \Delta u^2_j}{(u_j)^2}.
\label{vecpo-212}
\eeqs
Hence, the AB-potential term with flux $\alpha$ in $(x, y)$-space is changed
into the same form with flux $2 \alpha$ in $(u^1, u^2)$-space. It is 
worthwhile to note that the {\it Levi-Civit\`{a}} transformation is a mapping
from flat $(x, y)$-space to flat $(u^1, u^2)$-space unlike KS transformation
whose image space has a torsion and curvature\cite{kle95,kle96,park97}.
Therefore we can change the measure of Eq.(2.8) as \cite{kle95}
\beq
\prod_{j=2}^{N+1} \left( \frac{M}{2 \pi i \epsilon_s r_{j-1}} d \Delta 
 \vec{x}_j \right) \Rightarrow \prod_{j=1}^N \left( \frac{M}{2 \pi i \epsilon_s r_j}
 d \vec{x}_j \right).
\label{measure-213}
\eeq

>From Eqs. (2.11), (2.12), and (2.13)
$K[\vec{x}_b, \vec{x}_a ; E]$ is changed into 
\beq
K[\vec{x}_b, \vec{x}_a ; E] 
=(N+1) \int_0^{\infty} d \epsilon_s 
         \left( \frac{M}{2 \pi i \epsilon_s} \right)
         \int \left[ \prod_{j=1}^N \frac{4M}{2 \pi i \epsilon_s} d^2 \vec{u}_j
                                       \right]
        e^{i S_E^N[\vec{u}, \vec{u}']}
\label{fixeden-214}
\eeq
where
\beq
S_E^N[\vec{u}, \vec{u}'] = -\xi s + \sum_{j=1}^{N+1}
 \left[ 2M \frac{(\vec{u}_j - \vec{u}_{j-1})^2}{\epsilon_s} + 2 e \vec{A}^u_j
 \cdot \Delta \vec{u}_j + \epsilon_s E \vec{u}_j^2
                                                      \right]
\label{action-215}
\eeq
and 
\beq
\vec{A}^u = \frac{\alpha}{e} \frac{(u^2, -u^1)}{(u^1)^2 + (u^2)^2}.
\label{vecpo-216}
\eeq
The time-sliced action (\ref{action-215}) represents the AB plus 
harmonic oscillator system
whose mass and flux parameter are $4M$ and $2 \alpha$ respectively, and angular
frequency is 
\beq
\omega^2 = - \frac{E}{2M}.
\label{angfr-217}
\eeq

The well-known propagator\cite{park95,peak69} of the AB plus harmonic
oscillator system can be used to calculate the fixed-energy amplitude directly.
After treating the square root property of the {\it Levi-Civit\`{a}}
transformation carefully that was explained nicely in chapter 13 of Ref.[19],
$K[\vec{x}_b, \vec{x}_a ; E]$ becomes
\beq
K[\vec{x}_b, \vec{x}_a ; E] = \frac{1}{4} \int_0^{\infty} ds
\left( K[\vec{u}_b, \vec{u}_a ; s] + K[-\vec{u}_b, \vec{u}_a ; s] \right]
\label{fixeden-218}
\eeq
where
\beq
K[\vec{u}_b, \vec{u}_a ; s] = \sum_m e^{i m (\phi_b - \phi_a)}
K_m[u_b, u_a ; s].
\label{fixeden-219}
\eeq
Here, $\phi$ is defined as polar angle in $(u^1, u^2)$-space
\beqs
u^1&=& u \cos \phi \\ \nn
u^2&=& u \sin \phi
\label{polar-220}
\eeqs
and 
\beqs
K_m[u_b, u_a ; s] 
&=& \frac{(4M) \omega}{2 \pi i \sin \omega s } e^{-i \xi s} \\  \nn
&\times&    exp \left[ \frac{i (4M) \omega}{2} (u_a^2 + u_b^2) \cot \omega s 
                                                                    \right]
    I_{\mid m + 2 \alpha \mid} 
    \left( \frac{-i (4M) \omega u_a u_b}{\sin \omega s} \right).
\label{fixeden-221}
\eeqs
After inserting Eqs. (2.19) and (2.21) into 
Eq.(2.18) and representing the result in terms of the polar 
coordinate in $(x, y)$ space: $\theta = 2 \phi$ and $r = u^2$, one 
arrives at the final form
of fixed-energy amplitude  
\beq
K[\vec{x}_b, \vec{x}_a ; E] = \sum_{-\infty}^{\infty}
 e^{i m (\theta_b - \theta_a)} K_m[x_b, x_a ;E]
\label{fixeden-222}
\eeq
where
\beqs
& &K_m[x_b, x_a ;E]  \\ \nn
&=&\frac{M \omega}{\pi i} \int_0^{\infty} ds \frac{e^{-i \xi s}}{\sin \omega s}
   exp \left[ 2 i M \omega (x_a + x_b) \cot \omega s \right]
   I_{2 \mid m + \alpha \mid} \left( \frac{-4 i M \omega}{\sin \omega s}
                                     \sqrt{x_a x_b}  \right).
\label{fixeden-223}
\eeqs
$K_m[x_b, x_a ;E]$ is a fixed-energy amplitude associated to sum over all
possible paths within the $m^{th}$ homotopy class.

In order to obtain the energy spectrum one has to check the poles of 
fixed-energy amplitude carefully. This is easily achieved 
by changing the variable
$v = - i / \sin \omega s$ and performing the $v$-integration explicitly
using the integral formula\cite{gra65}
\beqs
& &\int_0^{\infty} \frac{dx}{\sqrt{x^2 + z^2}}
\left( \frac{\sqrt{x^2 + z^2} \pm z}{x} \right)^{\mu}
exp [-p \sqrt{x^2 + z^2}] I_{\nu}(c x)  \\  \nn
&=& \frac{1}{cz} \frac{\Gamma \left( \frac{1 + \nu \pm \mu}{2} \right)}
                      {\Gamma \left(1 + \nu \right) }
    W_{\pm \frac{\mu}{2}, \frac{\nu}{2}}(z_+)
    M_{\pm \frac{\mu}{2}, \frac{\nu}{2}}(z_-)
\label{intfo-224}
\eeqs
where $W_{a, b}$ and $M_{a, b}$ are usual Whittaker functions and 
$z_{\pm} = z ( p \pm \sqrt{p^2 - c^2} )$.

The $v$-integration makes $K_m[x_b, x_a ; E]$ to be 
\beqs
K_m[x_b, x_a ; E]  
&=& \frac{1}{4 \pi i \omega \sqrt{x_a x_b}}
    \frac{ \Gamma \left( \frac{1}{2} + \mid m + \alpha \mid + \frac{\xi}
                          {2 \omega} \right) }
         { \Gamma \left( 1 + 2 \mid m + \alpha \mid \right) } \\ \nn
&\times&    W_{-\frac{\xi}{2 \omega}, \mid m + \alpha \mid} 
                                 \left( 4 M \omega Max(x_a, x_b) \right) 
   M_{-\frac{\xi}{2 \omega}, \mid m + \alpha \mid}
                                 \left( 4 M \omega Min(x_a, x_b) \right).
\label{fixeden-225}
\eeqs
The energy spectrum of the ABC system is deduced from the poles of the 
gamma function in the numerator:
\beq
E_{n, m} = -\frac{1}{2} \frac{M \xi^2}{ (n + \mid m + \alpha \mid - 
                                        \frac{1}{2})^2}.
\hspace{1cm} n = 1, 2, \cdots 
\label{eigval-226}
\eeq
This spectrum is in agreement with the result corresponding to the regular
solution obtained in Ref.[12].

In Sec.III The main idea of Ref.[14] for the evaluation 
of propagator in spin-1/2 ABC system will be briefly reviewed.
\newpage

\section{Proper incorporation of the self-adjoint extension method into the
Green's function formalism}                                                                            
In this section we will discuss how to incorporate the self-adjoint extension 
method into the Green's function formalism briefly.  
A one-dimensional Hamiltonian below will be a good model for this purpose:
\beq
H = H_V + v \delta (x)
\label{hamil-31}
\eeq
where $H_V$ is 
\beq
H_V = \frac{p^2}{2} + V(x).
\label{hamil-32}
\eeq

Taking a Laplace transform to the well-known integral equation
\beq
G[x_1, x_2 ; t] 
= G_V[x_1, x_2 ; t] - v \int_0^t ds \int dx 
G_V[x_1, x, t-s] \delta (x) G[x, x_2 ; s]
\label{inteq-33}
\eeq
where $G[x_1, x_2 ; t]$ and $G_V[x_1, x_2 ; t]$ are euclidean propagators of 
$H$ and $H_V$ respectively, it is straightforward to derive the relation
\beq
\hat{G}[x_1, x_2 ; E]  
= \hat{G}_V[x_1, x_2 ; E] - v \hat{G}_V[x_1, 0 ; E] \hat{G}[0, x_2 ; E].
\label{geq-34}
\eeq
>From Eq.(3.4) one can easily obtain 
\beq
\hat{G}[0, x_2; E] = \frac{\hat{G}_V[0, x_2; E]}
                          {1 + v \hat{G}_V[0, 0, E]}
\label{geq-35}
\eeq
by taking a $x_1 \rightarrow 0$ limit. Upon combining Eqs. (3.4)
and (3.5) energy-dependent Green's function $\hat{G}[x_1, x_2; E]$
is easily calculated from $\hat{G}_V[x_1, x_2; E]$:
\beq
\hat{G}[x_1, x_2; E]  
=\hat{G}_V[x_1, x_2; E] - 
\frac{\hat{G}_V[x_1, 0, E] \hat{G}_V[0, x_2; E]}
     {\frac{1}{v} + \hat{G}_V[0, 0; E]}.
\label{geq-36}
\eeq
Applying the above method to the $V(x) = 0$ case which was done firstly
in Ref.[24] produces the energy-dependent Green's function for the 
$\delta$-function potential system
\beq
\hat{G}[x, y, E]  
= \frac{e^{-\sqrt{2E} \mid x-y \mid}}{\sqrt{2 E}}
   - v \frac{e^{-\sqrt{2E} ( \mid x \mid + \mid y \mid )}}
            {\sqrt{2E} ( \sqrt{2E} + v )}
\label{geq-37}
\eeq
and euclidean time-dependent propagator by taking an inverse Laplace transform
to Eq.(3.7)
\beq
     G[x, y; t] = G_{0}[x, y;t] - v\int_{0}^{\infty}dz
     e^{-vz} G_{0}[\mid x \mid, -\mid y \mid - \mid z
     \mid;t],
\label{geq-38}
\eeq
where $G_0[x, y;t]$ is the time-dependent propagator for a 
one-dimensional free particle.

It is a simple matter to show that $G[x, y, t]$ satisfies the well-known 
boundary condition in the Kronig-Penney model
\beq
    \dd{G}{x}(0^{+}, y;t) - \dd{G}{x}(0^{-}, y;t)
    = 2 v G[0, y;t].
\label{geq-39}
\eeq
An important observation that there are two different ways
to calculate the bound-state spectrum from energy-dependent Green's
function
will play a crucial role for the 
calculation of higher dimensional cases.  
One of the two ways to obtain the bound-state energy is checking the poles of 
energy-dependent Green's function, which is an universal property of 
$\hat{G}[x, y, E]$.
The other is applying the boundary 
condition (\ref{geq-39}) to the energy-dependent Green's function:  
\beq
     \dd{ \hat{A} }{x}[0^{+}, y;E] - \dd{ \hat{A} }{x}[0^{-}, y;
     E] = 2 v \hat{A} [0, y;E]
\label{geq-310}
\eeq
where $\hat{A}[x, y, E] \equiv \hat{G}[x, y, E] - \hat{G}_0[x, y, E]$.
This means that the boundary condition (\ref{geq-39}) plays an important
role for the occurrence of a bound state. In Ref.[14] the use of these two 
different ways is examplified to derive the relation between
the self-adjoint extension parameter and the bare coupling constant in 
two- and three-dimensional systems. Once this relation is obtained, 
the derivation of energy-dependent Green's function is straightforward.

The above-mentioned method, however, cannot be applied directly to the 
spin-1/2 ABC system because the energy-dependent Green's  function 
of a spinless ABC system, which can be obtained from fixed-energy amplitude
(2.22) and (2.25) through the relation
\beq
\hat{G}[\vec{x}_b, \vec{x}_a; E] = i K[\vec{x}_b, \vec{x}_a; -E],
\label{geq-311}
\eeq
diverges when either $\vec{x}_a$ or $\vec{x}_b$ approaches zero.
This is because of the magnetic flux tube located at the origin. 
In such a case as suggested in Ref.[14] Eq.(3.6) must be modified
to be
\beq
\hat{G}[\vec{x}_b, \vec{x}_a; E]  
=\hat{G}_V[\vec{x}_b, \vec{x}_a; E] - 
\frac{ \hat{G}_V[\vec{x}_b, \vec{\epsilon}_1; E] 
       \hat{G}_V[\vec{\epsilon}_2, \vec{x}_a; E] }
     {\frac{1}{v} + \lim_{\epsilon_2 \rightarrow \epsilon_1^+}
       \hat{G}_V[\vec{\epsilon}_2, \vec{\epsilon}_1; E]}.
\label{geq-312}
\eeq
In Eq.(3.12) the limit $\epsilon_1 \rightarrow 0$ should be taken
at the final stage of calculation.

It is shown in Ref.[14] through the explicit calculation that 
this method produces a physically relevant 
propagator consistent with the boundary condition deduced by self-adjoint
extension method and derives the relationship between the self-adjoint 
extension parameter and the coupling constant consistently and 
convincingly.

In next section we will apply this method to the spin-1/2 ABC system 
and obtain the energy-dependent Green's function explicitly.

\newpage

\section{Propagator for spin-1/2 ABC system}

>From Eqs.(2.22) and (2.25) the energy-dependent 
Green's function $\hat{G}_B[\vec{x}_b, \vec{x}_a; E]$ of the spinless ABC
system is 
\beq
\hat{G}_B[\vec{x}_b, \vec{x}_a; E] = \sum_{m= -\infty}^{\infty} 
  e^{i m (\theta_b - \theta_a)} \hat{G}_m^B[[x_b, x_a; E]
\label{engre-41}
\eeq
where
\beqs
\hat{G}_m^B[[x_b, x_a; E]  
&=& \frac{1}{4 \pi \Omega \sqrt{x_a x_b}}
    \frac{\Gamma \left( \frac{1}{2} + \mid m + \alpha \mid + 
                        \frac{\xi}{2 \Omega} \right) }
         {\Gamma ( 1 + 2 \mid m + \alpha \mid ) }  \\  \nn
&\times& W_{-\frac{\xi}{2 \Omega}, \mid m + \alpha \mid} ( 4 M \Omega Max(x_a, x_b))
    M_{-\frac{\xi}{2 \Omega}, \mid m + \alpha \mid} ( 4 M \Omega Min(x_a, x_b))
\label{engre-42}
\eeqs
and 
\beq
\Omega^2 = \frac{E}{2 M}.
\label{angfre-43}
\eeq

When spin degree of freedom is added to the ABC system, it is well known
[7, 9, 13] that a delta-function potential appears 
because of the Zeeman interaction.
The Hamiltonian, therefore, for a large component of Dirac spinor is 
\beq
H_F = H_B + v \delta (\vec{x})
\label{hamil-44}
\eeq
where $H_B$ is given in Eq.(\ref{hamil-21}).

Now let us define the energy dependent Green's function of spin-1/2
system in terms of each homotopy classes: 
\beq
\hat{G}^F[\vec{x}_b, \vec{x}_a; E] = \sum_{m = -\infty}^{\infty}
e^{i m (\theta_b - \theta_a)} \hat{G}_m^F[x_b, x_a; E].
\label{engre-45}
\eeq

If one uses the asymptotic formula of the modified Bessel function
$ I_{\nu}(z) \propto z^{\nu} $ in Eq.(2.23), one can realize 
easily that the application of the self-adjoint extension method must
be restricted to the domain $\mid m + \alpha \mid < 1/2$ because of the 
normalizability condition. 
This is also conjectured in Ref.[25].
We, therefore, confine our attention to this
domain for the time being. 

Upon combining Eqs.(3.12), (4.1) and (4.5)
one can get the following relation:
\beqs
& &\hat{A}_m[x_b, x_a; E]  \\  \nn
&=& f_m(\epsilon_1, \epsilon_2; E)
    \frac{W_{-\frac{\xi}{2 \Omega}, \mid m + \alpha \mid}(4 M \Omega x_a)}
         {\sqrt{x_a}}
    \frac{W_{-\frac{\xi}{2 \Omega}, \mid m + \alpha \mid}(4 M \Omega x_b)}
         {\sqrt{x_b}}
\label{engre-46}
\eeqs
where
\beq
\hat{A}_m[x_b, x_a; E]  
\equiv \hat{G}_m^F[x_b, x_a; E] - \hat{G}_m^B[x_b, x_a; E]
\label{engre-47}
\eeq 
and 
\beqs
f_m(\epsilon_1, \epsilon_2; E)  
&=&- \frac{1}{\frac{1}{v} + \hat{G}_m^B[\epsilon_2, \epsilon_1; E]}
     \frac{1}{\sqrt{\epsilon_1 \epsilon_2} (4 \pi \Omega)^2}
     \left( 
            \frac{\Gamma \left( \frac{1 + 2 \mid m + \alpha \mid + 
                                      \frac{\xi}{\Omega}}{2} \right)}
                 {\Gamma (1 + 2 \mid m + \alpha \mid )}
                                                       \right)^2 \\ \nn
&\times& M_{- \frac{\xi}{2 \Omega}, \mid m + \alpha  \mid} (4M \Omega \epsilon_1)
   M_{- \frac{\xi}{2 \Omega}, \mid m + \alpha  \mid} (4M \Omega \epsilon_2).
\label{engre-48}
\eeqs

In order to apply the boundary condition at the origin deduced from the 
self-adjoint extension method, it is more convenient to express
$\hat{A}_m[x_b, x_a; E]$ in terms of hypergeometric function:
\beq
\hat{A}_m[x_b, x_a; E] = f_m(\epsilon_1, \epsilon_2; E)
                         g_m(x_a)  g_m(x_b)
\label{engre-49}
\eeq
where
\beqs
g_m(r)  
&=& (4 M \Omega)^{\frac{1}{2} + \mid m + \alpha \mid} e^{-2 M \Omega r}
    r^{\mid m + \alpha \mid}  \\  \nn
&\times&    U \left( \frac{1}{2} + \mid m + \alpha \mid + \frac{\xi}{2 \Omega},
             1 + 2 \mid m + \alpha \mid; 4 M \Omega r
                                                         \right)
\label{engre-410}
\eeqs
and
\beq
U(a, b, z)  
= \frac{\pi}{\sin \pi b}
   \left[ \frac{F(a, b, z)}{\Gamma (1+a-b) \Gamma(b)} - z^{1 - b}
          \frac{F(1+a-b, 2-b, z)}{\Gamma (a) \Gamma (2-b)}
                                                          \right].
\label{engre-411}
\eeq
Now let us apply the boundary condition which is obtained from the 
self-adjoint extension method
\beqs
& &\lim_{x_a \rightarrow 0} x_a^{\mid m + \alpha \mid} \hat{A}_m[x_b, x_a; E] \\ \nn
&=&\lambda_m \lim_{x_a \rightarrow 0}
   \frac{1}{x_a^{\mid m + \alpha \mid}}
   \left[ \hat{A}_m[x_b, x_a; E] - \frac{1}{x_a^{\mid m  + \alpha \mid}}
         \left( \lim_{x_a' \rightarrow 0} x_a'^{\mid m + \alpha \mid}
              \hat{A}_m[x_b, x_a'; E]   \right)
                                                 \right]
\label{engre-412}
\eeqs 
where $\lambda_m$ is a self-adjoint extension parameter.  
The boundary condition (4.12)
is derived by one of us in Ref.[25].

After using the asymptotic formula of $U(a, b, z)$ and performing a tedious
calculation, one can show that 
the boundary condition (4.12) yields the following relation
\beq
\frac{\Gamma \left(\frac{1}{2} + \mid m + \alpha \mid + \frac{\xi}{2 \Omega} \right)}
     {\Gamma \left(\frac{1}{2} - \mid m + \alpha \mid + \frac{\xi}{2 \Omega} \right)} = 
\frac{1}{\lambda_m (4 M \Omega)^{2 \mid m + \alpha \mid}}
 \frac{\Gamma (2 \mid m + \alpha \mid )}
      {\Gamma (-2 \mid m + \alpha \mid )}.
\label{rel-413}
\eeq

Hence, by solving Eq.(4.13) one can obtain the bound-state energy.
Although Eq.(\ref{rel-413}) is too complicated to evaluate the bound-state
energy explicitly, $\lambda_m \rightarrow 0$ and $\infty$ limiting 
features are interesting. In these cases the bound-state spectra are explicitly
determined as poles of the gamma function,i.e.,
\beqs
E_{n,m}&=& \frac{1}{2} \frac{M \xi^2}
                           {(n - \frac{1}{2} + \mid m + \alpha \mid)^2}
\hspace{1.0cm} (\lambda_m = 0)  \\  \nn
E_{n,m}&=& \frac{1}{2} \frac{M \xi^2}
                           {(n - \frac{1}{2} - \mid m + \alpha \mid)^2}
\hspace{1.0cm} (\lambda_m = \infty).  \\  \nn
& &\hspace{5.0cm} n = 1, 2, \cdots
\label{eneig-414}
\eeqs

These bound-state energies coincide exactly with those of regular and singular
solutions given in Ref.[12]. The absence of minus sign is due to the
euclidean characteristic.

Now the relation between the self-adjoint extension parameter
$\lambda_m$
and the bare coupling constant $v$ can be explored. 
This relation is easily obtained through the comparision of Eq.(4.13) with 
the poles of $\hat{A}_m[x_b, x_a, E]$
\beq
\frac{1}{v} + \hat{G}_m^B[\epsilon_2, \epsilon_1; E] = 0.
\label{pole-1}
\eeq
Counting on the asymptotic formula of 
$U(a, b, z)$ one can show that  the poles of $\hat{A}_m[x_b, x_a; E]$ arise
when the following relation is satisfied: 
\beqs
\frac{\Gamma \left(\frac{1}{2} + \mid m + \alpha \mid + \frac{\xi}{2 \Omega} 
\right) }
     {\Gamma \left(\frac{1}{2} - \mid m + \alpha \mid + \frac{\xi}{2 \Omega} 
                                              \right) } &=&
        - \frac{1}{(4 M \Omega)^{2 \mid m + \alpha \mid}}
          \frac{\Gamma (2 \mid m + \alpha \mid )}
               {\Gamma (-2 \mid m + \alpha \mid )} \\  \nn
&\times& \frac{1}{(\epsilon_1 \epsilon_2)^{\mid m + \alpha \mid}}
    \left[ \frac{2 \pi \mid m + \alpha \mid}{M v} + 
           \left( \frac{\epsilon_1}{\epsilon_2} \right)^{\mid m + \alpha \mid}
                                                               \right].
\eeqs
Comparision of Eq.(4.16) with Eq.(4.13) enables one to get the relation
between the self-adjoint extension parameter and the coupling constant
\beq
\frac{1}{v} = - \frac{M}{2 \pi \mid m + \alpha \mid}
              \left[ \frac{(\epsilon_1 \epsilon_2)^{\mid m + \alpha \mid}}
                          {\lambda_m}
                    + \left( \frac{\epsilon_1}{\epsilon_2}  
                                           \right)^{\mid m + \alpha \mid}
                                                           \right].
\label{eneig-415}
\eeq
This relation makes the denominator of Eq.(4.8) to be
\beqs
\frac{1}{v} + \hat{G}_m^B[\epsilon_2, \epsilon_1; E]  
&=& - \frac{M}{2 \pi \mid m + \alpha \mid} (\epsilon_1 \epsilon_2)^{\mid m + \alpha \mid} \\ \nn
&\times& \left[ \frac{1}{\lambda_m} - (4 M \Omega)^{2 \mid m + \alpha \mid}
       \frac{\Gamma (-2 \mid m + \alpha \mid) \Gamma(\frac{1}{2} + \mid m + 
               \alpha \mid + \frac{\xi}{2 \Omega} )}
            {\Gamma(2 \mid m + \alpha \mid) \Gamma (\frac{1}{2} - \mid m + 
               \alpha \mid + \frac{\xi}{2 \Omega} )}
                                                       \right].
\label{eneig-416}
\eeqs
By inserting Eq.(4.18) into (4.8) and taking a limit
$\epsilon_2 \rightarrow \epsilon_1^+$ and $\epsilon_1 \rightarrow 0$, $f_m(\epsilon_1, \epsilon_2;
E)$ becomes the $\epsilon_1$- and $\epsilon_2$-independent finite quantity
\beqs
& &\lim_{\epsilon_1, \epsilon_2 \rightarrow 0}  
   f_m(\epsilon_1, \epsilon_2; E)\\  \nn
&=& \frac{2 \pi \mid m + \alpha \mid}{M}
    \frac{(4 M \Omega)^{1 + 2 \mid m + \alpha \mid}}{(4 \pi \Omega)^2}
    \left( \frac{\Gamma(\frac{1}{2} + \mid m + \alpha \mid + 
                        \frac{\xi}{2 \Omega} )}
                {\Gamma (1 + 2 \mid m + \alpha \mid) }
                                                      \right)^2  \\  \nn
&\times& \left[
               \frac{1}{\lambda_m} - (4 M \Omega)^{2 \mid m + \alpha \mid}
               \frac{\Gamma(-2 \mid m + \alpha \mid)
                     \Gamma(\frac{1}{2} + \mid m + \alpha \mid +
                            \frac{\xi}{2 \Omega} )}
                    {\Gamma(2 \mid m + \alpha \mid)
                     \Gamma(\frac{1}{2} - \mid m + \alpha \mid +
                            \frac{\xi}{2 \Omega} )}
                                                      \right]^{-1}.
\label{eneig-417}
\eeqs

Substituting Eq.(4.19) into Eq.(4.6) directly  $\hat{A}_m[x_b, x_a; E]$ 
is easily obtained. 
Hence, the energy-dependent Green's function for the spin-1/2 system is 
\beqs
\hat{G}^F[\vec{x}_b, \vec{x}_a; E] 
&=& \sum_{\mid m + \alpha \mid > 1/2}
   e^{im(\theta_b - \theta_a)} \hat{G}_m^B[x_b, x_a; E]  \\  \nn
&+& \sum_{\mid m + \alpha \mid < 1/2}
   e^{im(\theta_b - \theta_a)}
   \Bigg[ 
         \hat{G}_m^B[x_b, x_a; E]  \\  \nn
& &\hspace{1.5cm}         
+ \frac{2 \pi \mid m + \alpha \mid}{M} 
         \frac{(4 M \Omega)^{1 + 2 \mid m + \alpha \mid}}{(4 \pi \Omega)^2}
         \left( \frac{\Gamma (\frac{1}{2} + \mid m + \alpha \mid + 
                              \frac{\xi}{2 \Omega} )}
                     {\Gamma (1 + 2 \mid m + \alpha \mid)}
                                                          \right)^2 \\ \nn
& &\hspace{1.5cm}   
\times \left[ \frac{1}{\lambda_m} - (4 M \Omega)^{2 \mid m + \alpha \mid}
                \frac{\Gamma (-2 \mid m + \alpha \mid)
                      \Gamma (\frac{1}{2} + \mid m + \alpha \mid
                              + \frac{\xi}{2 \Omega})}
                     {\Gamma (2\mid m + \alpha \mid)
                     \Gamma (\frac{1}{2} - \mid m + \alpha \mid
                              + \frac{\xi}{2 \Omega})}
                                                        \right]^{-1} \\  \nn
& &\hspace{1.5cm}         
\times \frac{W_{-\frac{\xi}{2 \Omega}, \mid m + \alpha \mid}
                          (4 M \Omega x_a)}  {\sqrt{x_a}}
                  \frac{W_{-\frac{\xi}{2 \Omega}, \mid m + \alpha \mid}
                          (4 M \Omega x_b)}  {\sqrt{x_b}}
                                                            \Bigg].
\label{eneig-418}
\eeqs
Of course we can obtain the fixed-energy amplitude from Eq.(4.20) by using the 
relation (3.11).
\newpage

\section{Conclusion}

We derived the propagators of the spinless and spin-1/2 ABC systems explicitly. 
For the derivation of propagator in the spinless ABC system we applied the 
Duru-Kleinert method which was used firstly 
to evaluate the propagator
for the case of 
hydrogen atom. It is found that the spinless ABC system with mass $M$ and 
flux parameter $\alpha$ is reduced to the AB plus harmonic oscillator
system with mass $4M$, flux parameter $2 \alpha$ and angular frequency
$\sqrt{-E/2M}$ through the Levi-Civit\`{a} transformation that is a 
two-dimensional counterpart of the KS transformation. The fact that 
the Levi-Civit\`{a} transformation is mapping from flat space to flat space
unlike the KS transformation makes the derivation of propagator in 
the spinless ABC system extremely simple. The final form of the propagator 
is expressed as an winding number representation and the bound-state
spectrum deduced from it is in agreement with that corresponding to the 
regular solution at the origin suggested in the previous article.

With the propagator of the spinless ABC system we analyzed also the 
spin-1/2 ABC system. For the derivation of the propagator in the spin-1/2
ABC system 
the following two different approaches were used to get the 
bound-state spectrum: \\
1. by checking the poles of the energy-dependent Green's function. \\
2. by requiring the boundary condition derived from the self-adjoint 
extension to the  
energy-dependent Green's function.  \\
Identification of these two different spectra naturally 
leads one to the relation between the
self-adjoint extension parameter and the coupling constant.
Substitution of this relation into the denominator of the energy-dependent
Green's function enabled us to obtain the finite 
propagator of the spin-1/2 ABC system.

Since AB interaction is acted on charge and magnetic flux tube, 
interaction between anyons requires a Coulomb modification. Although 
the expression of the two-dimensional Coulomb term may be disputable, 
reflecting from the fact that real world is three-dimensional one and 
two dimension is embedded in it, the propagator obtained
in the present paper might be used for the study of the time-dependent
scattering or statistical properties of anyon system. 


\end{document}